# ARRIVAL FLOW PROFILE ESTIMATION AND PREDICATION FOR URBAN ARTERIALS USING LICENSE PLATE RECOGNITION DATA

**Hao WU,** Postdoctoral Fellow
Department of Electrical and Electronic Engineering, the Hong Kong Polytechnic University
Hung Hom Kowloon, Hong Kong
E-mail: hao96.wu@polyu.edu.hk

**Jiarong YAO,** Research Fellow
School of Electrical and Electronic Engineering, Nanyang Technological University
50 Nanyang Avenue, Singapore, 639798
E-mail: yaojoanna2018@gmail.com

**Peize KANG,** Ph.D. Candidate
Key Laboratory of Road and Traffic Engineering of the Ministry of Education
College of Transportation Engineering, Tongji University, Shanghai, China 201804
E-mail: kangpeizhe@tongji.edu.cn

**Chaopeng TAN,** Postdoctoral Fellow
Department of Transport and Planning, Delft University of Technology
Gebouw 23, Stevinweg 1, Delft, 2628 CN, Netherlands
E-mail: tantantan951122@gmail.com

**Yang CAI,** Engineering
Shanghai Urban Construction Design and Research Institute (Group) Co., Ltd, Shanghai, China
E-mail: caiyang@sucdri.com

**Junjie ZHOU,** Business Analyst
Beijing Kuaishou Technology Co., Ltd, Beijing, China
E-mail: 1902402944@qq.com

**Edward CHUNG,** Professor
Department of Electrical and Electronic Engineering, the Hong Kong Polytechnic University
Hung Hom Kowloon, Hong Kong
E-mail: edward.cs.chung@polyu.edu.hk

**Keshuang TANG\*,** Corresponding Author, Professor
Key Laboratory of Road and Traffic Engineering of the Ministry of Education
College of Transportation Engineering, Tongji University, Shanghai, China 201804
E-mail: tang@tongji.edu.cn

Word count: 8,369 words text + 13 tables/figures

February, 2025

**ABSTRACT**

Arrival flow profiles enable precise assessment of urban arterial dynamics, aiding signal control optimization. License Plate Recognition (LPR) data, with its comprehensive coverage and event-based detection, is promising for reconstructing arrival flow profiles. This paper introduces an arrival flow profile estimation and prediction method for urban arterials using LPR data. Unlike conventional methods that assume traffic homogeneity and overlook detailed traffic wave features and signal timing impacts, our approach employs a time partition algorithm and platoon dispersion model to calculate arrival flow, considering traffic variations and driving behaviors using only boundary data. Shockwave theory quantifies the piecewise function between arrival flow and profile. We derive the relationship between arrival flow profiles and traffic dissipation at downstream intersections, enabling recursive calculations for all intersections. This approach allows prediction of arrival flow profiles under any signal timing schemes. Validation through simulation and empirical cases demonstrates promising performance and robustness under various conditions.

# I. INTRODUCTION

Urban arterial functions as the primary channels facilitating the movement of medium and long-distance urban traffic, thereby underscoring the significance of enhancing the surveillance and control of traffic operations on these urban arterial routes. Profiling the arrival flow enables the precise assessment of the spatial-temporal dynamics within each road segment, thereby yielding a critical insight into the traffic state of the urban arterial. Consequently, the reconstruction of arrival flow profiles for urban arterials emerges as a pivotal focal point within the realm of urban traffic management and optimization.

Arrival flow profiles are extensively studied in on-ramp metering [1], [2], while their emergence in urban signal control research dates back to 2005 with Nikolaos credited as a pioneer [3]. However, direct efforts in estimating arrival flow profiles remain limited, with the focus shifting towards indirect estimation approaches. These approaches fall into two main categories: input-output models [4], [5], [6], [7] and shockwave-based models [8], [9], [10], [11], [12]. The former dynamically reconstructs traffic flow by generating cumulative arrival and departure curves at signalized intersections, while it relies on full-sampled data sources and assumes free-flow travel time. The latter dynamically reconstructs traffic flow by studying the formation and dissipation of vehicular queues at signalized intersections, providing detailed insights into dynamic traffic flow, while only a minority of these studies consider driving diversity and traffic flow fluctuations. While some explore alternative theories [13], [14], most existing research primarily focuses on single road links, neglecting the influence of signal timing of the upstream intersections within the scale of an urban arterials. Meanwhile, given that the essence of arrival flow profile estimation lies in accurately reconstructing traffic flow, alternative methodologies can be adapted to indirectly indicate the traffic dynamics on urban arterials, for instance, platoon dispersion model, [15], [16], [17], trajectory reconstruction model, [18], [19], [20], [21], [22], [23], [24], [25] and [26].

Recently, License Plate Recognition (LPR) systems have seen rapid development and widespread adoption in many cities in China. An example of this is the LPR system in Shanghai, which uses over 42,500 LPR cameras [27]. LPR data could record the registered vehicles' plate number and passing time and provide information about vehicles' operating situations. Besides, multiple LPR cameras could be combined to capture the movement of vehicles between intersections and could thus obtain the travel time information. As a result, LPR data are highly competitive for estimating arrival flow profiles for two key reasons: LPR data's ability to detect individual vehicle information makes it a preferable substitute of other fixed-location data sources with similar characteristics, such as high-resolution radar data and Radio-Frequency Identification Data (RFID); Besides, with vehicle IDs and license plate information, this methodology also holds potential for analyzing other types of mobile detector data.

Hence, we introduce an arrival flow profile estimation and prediction method using LPR data. Initially, input data is collected solely from LPR cameras at the entrance boundaries of the studied urban arterial. Utilizing known traffic flow information from these boundaries and considering traffic flow variations, we partition time into smaller segments and calculate the arrival flow for each time segment for these sections, employing a platoon dispersion model that considers diverse driving behaviors. Subsequently, a piecewise function is established through shockwave theory to characterize the correlation between arrival flow and profile. We then quantify the relationship between the arrival flow profile and its dissipated traffic flow toward downstream intersections by considering signal timing. This data serves as the arrival flow input for downstream intersections in profile estimation. This recursive process allows us to estimate the arrival flow profile for the entire arterial. Additionally, our manuscript presents a method for predicting arrival flow profiles for any given signal timing schemes. It's crucial to distinguish between arrival flow profile estimation, relevant to fixed signal timing and providing an estimation under current conditions, and arrival flow profile prediction, applicable to any given signal timing scenario.

This study represents the first endeavor to create highly detailed arrival flow profiles for urban arterials. Our key contributions include an improved approach compared to existing methods that assume traffic state homogeneity within cycles and overlook nuanced traffic wave features and signal timing impacts. Our proposed method captures intricate features like driving diversity, fluctuation in traffic state and signal timing intricacies, allowing for a precise assessment of urban arterial dynamics. This approach is crucial for signal control evaluation, representing a crucial step in advancing traffic management and optimization. Practically, our arrival flow profile prediction method serves as a valuable tool for evaluating signal timing efficacy on urban arterials before implementation by traffic authorities. This method serves as an effective alternative to traffic simulation tools, especially in situations where the availability of such tools is limited, showcasing broader applicability for signal control evaluation and optimization compared to simulation-driven approaches.

The remainder of this article is organized as follows: Section II conducts a comprehensive review of prior research in the fields related to arrival flow profile estimation and prediction, emphasizing the existing gaps in the literature. Subsequently, Section III and Section IV elucidate the methodologies employed for the estimation and prediction of the arrival flow profile for urban arterial, respectively. Section V then provides a detailed presentation of verification through an empirical case and a simulation case. Lastly, Section VI offers a conclusive discussion of the findings and outlines directions for future research.

## II. LITERATURE REVIEW

Arrival flow profile has been widely researched within the domain of on-ramp metering, [1], [2]. These studies seek to determine whether the demand at on-ramps surpasses the capacity of ramp metering systems, necessitating further control and management strategies. Holding equal significance in urban arterials, arrival flow profile serves as an invaluable tool for assessing signal control strategies and facilitating optimization of interrupted facilities. In the field of urban signal control, the research on arrival flow profiles traces its roots back to 2005, credited to Nikolaos et al [3] as the pioneer. Their pioneering research introduced an analytical method aimed at predicting platoon arrival profiles within urban arterials using loop detector data. Employing the principles of kinematic wave theory, this study proposed a Markov decision process model to predict arrival profiles. While limited follow-up studies exist on urban arterial arrival flow profiles, substantial efforts have been dedicated to dynamically modeling traffic flow in various arrival flow estimation methods. These dynamic traffic reconstruction methodologies, including arrival flow profiles and queue profiles, can be categorized into two groups: input-output model [4], [5], [6], [7], and shockwave-based model, [8], [9], [10], [11], [12].

In the context of the input-output model, traffic flow can be dynamically reconstructed by generating cumulative vehicle arrival and departure curves at signalized intersections. For example, Sharma [4] introduced a real-time estimation method for delay and maximum queue length based on input-output model, accurately recording arrival flow profiles through loop detector data. Although subsequent studies [5], [6] also focused on queue length estimation using arrival profile obtained by the input-output model, they often overlooked individual driving behavior. Later, Wu et al. [7] proposed an alternative method using LPR data, incorporating potential lane changing and overtaking behaviors for queue profile estimation. However, dynamic traffic flow reconstruction using the input-output model heavily relies on full-sampled data sources like loop detector data, which are limited by the location of traffic detectors and the quality of traffic data.

In the context of the shockwave model, traffic flow profiles can be accurately and dynamically reconstructed by estimating the formation and dissipation of shockwaves of vehicular queues at signalized intersections, utilizing various data sources. For example, Wu and Liu [8] introduced the Shockwave Profile Model (SPM) to simulate traffic dynamics using trajectory data. This model analytically derives the trajectories of shockwaves and divides road sections into small cells, allowing the reconstruction of queuing dynamics within each segment by tracking shockwave fronts. Subsequently, Mohsen and Nikolas [9] presented a queue profile estimation method that integrates dispersed probe vehicle data, shockwave theory, and data mining techniques. This approach captures queue evolutions at successive intersections and is applicable in oversaturated conditions. Subsequent research, similarly, based on mobile detector data, is exemplified, [10], [11]. In addition, Hu et al. [12] employed machine learning to model dynamic shockwave propagation and introduced a heuristic for measuring shockwave speed, thus enabling the estimation of queue profiles using data from connected vehicles.

While the shockwave model offers a more detailed insight into dynamic traffic flow, including the arrival and dissipation processes, only a minority of these studies consider driving diversity and traffic flow fluctuations. As traffic detectors and traffic state estimation methodologies continue to evolve, some researchers are exploring arrival flow profile reconstruction using alternative fundamental theories, such as probabilistic methods [13], [14]. However, these methods often require prior probabilities and parameter calibrations. Importantly, most of the aforementioned research is limited to single road links, where arrival flow profiles can be straightforwardly modeled based on the original traffic flow. In the context of urban arterials, arrival flow profiles at signalized intersections are not only influenced by traffic flow arrival pattern but also by signal timing, adding an additional layer of complexity.

Though research concerning the modeling of successive traffic flow on urban arterials is relatively limited, given that the essence of arrival flow profile estimation lies in accurately reconstructing traffic flow, alternative methodologies can be adapted to indirectly indicate the traffic dynamics on urban arterials. As a variant of the input-output method, two prominent approaches are the platoon dispersion model and the trajectory reconstruction model. The former establishes a relationship between upstream departures and downstream arrivals on each road link, enabling the estimation of downstream arrival rates based on upstream departure rates to generate the arrival flow profile [15], [16], [17]. The latter method employs emerging data sources to directly reconstruct vehicular trajectories along the entire urban arterial. By reconstructing trajectories, this model captures real-time arrival and dissipation processes, including the arrival flow profile, utilizing data from fixed-location detector data, [18], [19], [20], [21], or sample trajectories [22], [23], [24], [25], [26]. Thus, to overcome the limitations of previous studies utilizing input-output and shockwave models, this manuscript introduces an innovative method for estimating and predicting arrival flow profiles on urban arterials. This approach leverages License Plate Recognition (LPR) data and combines the strengths of both platoon dispersion and trajectory reconstruction models.

## III. ARRIVAL FLOW PROFILE ESTIMATION

In this manuscript, we introduce an innovative approach for estimating the arrival flow profile along urban arterial. The research scenario is illustrated in Fig. 1, where LPR cameras are strategically positioned along the boundaries of urban arterials. Consequently, the primary objective of this study revolves around formulating a methodology for inferring the arrival flow profile in sections devoid of LPR cameras. These sections have been categorized into two distinct groups: the "outer section", which encompasses areas with LPR cameras at the upstream intersection, enabling the detection of upstream traffic flow, and the "inner section", characterized by the absence of LPR cameras at the upstream intersection, resulting in its traffic flow being contingent upon the signal timing of the upstream intersection.

To estimate the arrival flow profile on both directions of the mainline of the urban arterial, we employ a step-by-step recursive calculation algorithm. As outlined in Fig. 1, taking the example of the north-bound direction of the mainline, the arrival flow profile of north-bound sections for intersections $j-1$, $j$, $j+1$, and $j+2$ should be estimated in sequence.

Starting with the outer section, which, in this case, is the north-bound section of intersection $j$, we calculate the vehicle inputs using direct traffic flow measurements from intersection $j-1$. These measurements are used in a platoon dispersion model to estimate the arrival traffic flow at the east-bound section of intersection $j$. Subsequently, we apply shockwave theory to estimate the arrival flow profile for this section. Furthermore, we calculate the actual dissipated traffic flow for this section during the research period, which also serves as the vehicle input for the next inner section (the north-bound section of intersection $j+1$). We repeat this recursive calculation process to estimate the arrival flow profile for all sections.

Thus, the subsequent discussion will delve into the four key steps: vehicle input processing, platoon dispersion estimation, shockwave reconstruction, and traffic volume estimation.

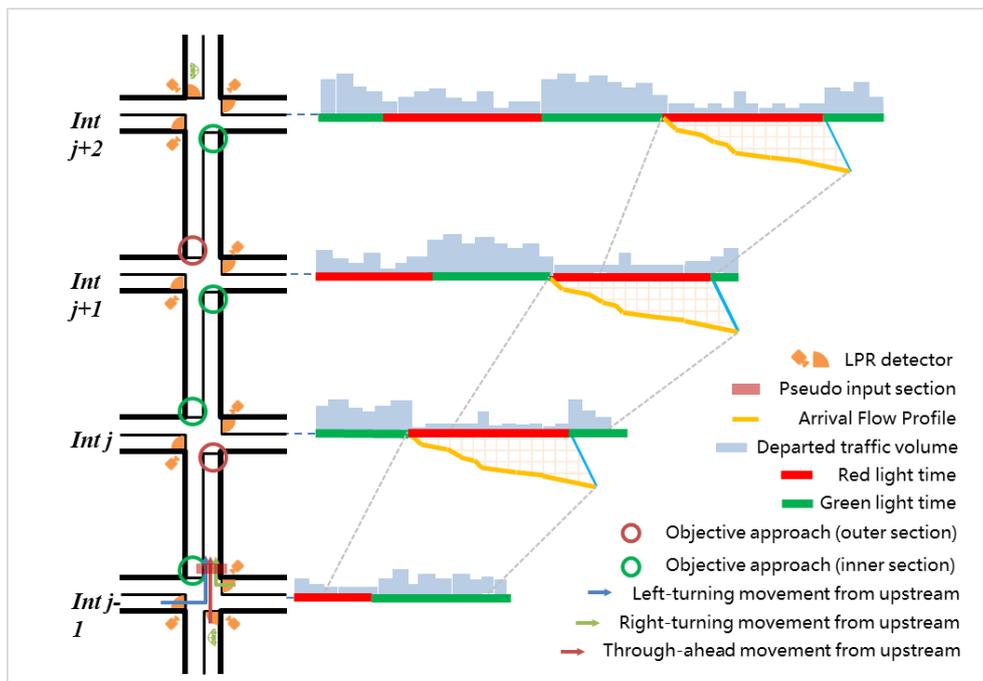

**Fig. 1.** Research scenario for arrival flow profile estimation

### A. Vehicle Input Processing

At the onset of our calculation, specifically in the outer section, input traffic volume is derived from three upstream movements, illustrated in Fig. 1. Considering the north-bound of intersection $j$ as our initial section, the three colored arrows represent vehicle inputs from diverse upstream traffic movements.

To simplify, the location of vehicle input is designated at the start location of road link, thereby creating a pseudo input section. Thus, given the precise recording of timestamps for each vehicle as it crosses the upstream stop-lines, the timestamp at which it passes through the pseudo input section can be further calculated (referred to as arrival timestamp).

$$t_{up}^{j,k,m} = t_{join}^{j-1,k,m} + \frac{x_{join}^{j-1,m}}{v_{inter}^{j-1}} \qquad (1)$$

Where $t_{up}^{j,k,m}$ is the arrival timestamp of vehicle $k$ from upstream movement $m$ of intersection $j$, here, $m$ can represent left-turning, through-ahead and right-turning movements, respectively; $t_{join}^{j-1,k,m}$ is the timestamp passing the stop-line of vehicle $k$ from movement $m$ of intersection $j-1$; $x_{join}^{j-1,m}$ is the distance from the stop-line of movement $m$ of intersection $j-1$, to the pseudo input section of intersection $j$; $v_{inter}^{j-1}$ is the speed under saturation flow rate, from movement $m$ of intersection $j-1$.

To transfer discrete vehicular arrival timestamps into arrival flow rates, the entire study period can be divided into multiple time intervals, denoted as $\Delta_t$. Subsequently, the arrival flow rate for each $\Delta_t$ interval can be computed. Initially, it is necessary to determine whether each vehicle passing through the pseudo input section belongs to a specific time interval.

$$\theta_{up}^{i,j,k,m} = \begin{cases} 1 \times \beta_{up}^k, & \text{when } t^i \le t_{up}^{j,k,m} < t^{i+1} \\ 0, & else \end{cases} \qquad (2)$$

Where $\theta_{up}^{i,j,k,m}$ represents a defined coefficient used to ascertain whether the arrival timestamp of vehicle $k$ from upstream movement $m$ at intersection $j$ falls within the time interval denoted as $[t^i, t^{i+1})$, and if this condition is met, $\theta_{up}^{i,j,k,m}$ is set to $\theta_{up}^{i,j,k,m} = 1 \times \beta_{up}^k$; otherwise, it is assigned a value of 0; $t^i$ denotes the start timestamp of time interval $i$, whose duration is $\Delta_t$; $\beta_{up}^k$ is the vehicle type conversion factor.

Subsequently, the summation of the coefficients $\theta_{up}^{i,j,k,m}$ for each vehicle allows us to calculate the total number of vehicles that have arrived during time interval $i$ from upstream movement $m$ at intersection $j$, denoted as $n_{up}^{i,j,m}$, as per Eq. (3). Furthermore, by aggregating the vehicles that have arrived during time interval $i$ from all upstream movements at intersection $j$, we can compute $n_{up}^{i,j}$ using Eq. (4).

$$n_{up}^{i,j,m} = \sum \theta_{up}^{i,j,k,m} \qquad (3)$$

$$n_{up}^{i,j} = \sum n_{up}^{i,j,m} \qquad (4)$$

## B. Estimation of Platoon Dispersion

Due to variations in driving behavior, vehicles exhibit dispersion as they approach the downstream stop-line. This phenomenon leads to the emergence of discrete traffic flow clusters, as depicted in Fig. 2.

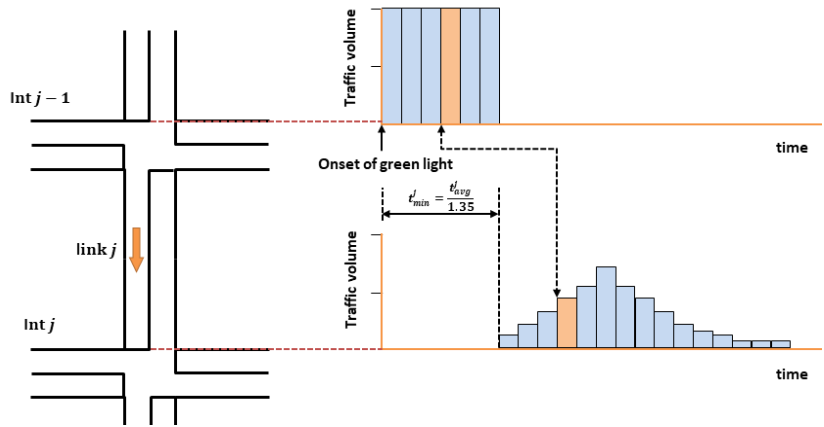

**Fig. 2.** Illustration of platoon dispersion

Regarding the input traffic volume, denoted as $n_{up}^{i,j}$, during time interval $i$, it should adhere to a distribution pattern in line with platoon dispersion theory. In this manuscript, Robertson's Platoon Dispersion Model is adopted due to its simplicity and efficiency [16]. Consequently, we employ a geometric distribution model to discretize the traffic flow $n_{up}^{i,j}$ along the road link, and Eq. (5) reveals that the downstream traffic flow represents a weighted average of historical traffic flows originating from the upstream pseudo input section.

$$n_{down}^{i,j} = \sum_{n=1}^{i-t_{min}^{j}} n_{up}^{i,j} \times F^j \times (1 - F^j)^{i-t_{min}^{j}-n} \tag{5}$$

Where $n_{down}^{i,j}$ is the downstream traffic volume during time interval $i$, originating from the pseudo input section of intersection $j$; $n_{up}^{i,j}$ is the traffic volume at the pseudo input section during time interval $i$ for intersection $j$; $t_{min}^{j}$ denotes the travel time of the fastest vehicle within the platoon on road link $j$; $F^j$ is the platoon dispersion coefficient for link $j$, used to quantify the level of platoon dispersion, and its calculation is outlined as follows:

$$F^j = \frac{1}{t_{avg}^{j} - t_{min}^{j} - 1} \tag{6}$$

Where $t_{avg}^{j}$ is the average travel time for platoons on road link $j$, which can be readily calibrated using LPR data; $t_{min}^{j}$ can be refined through calibration as $t_{min}^{j} = t_{avg}^{j}/1.35$.

## C. Shockwave Reconstruction

The discrete arrival data can be converted into a traffic volume distribution at the stop-line of the downstream intersection following platoon dispersion. Subsequently, we employ shockwave theory as the foundational framework, integrating discrete arrival data to achieve a precise estimate of the arrival flow profile.

In contrast to traditional traffic shockwave reconstruction using a static queuing wave velocity, we introduce the concept of a spatial-temporal zone, represented by shaded blue areas in Fig. 3. This zone outlines where traffic travels along the road link at an average speed $v_{lane}^{j}$ during a defined time interval $i$.

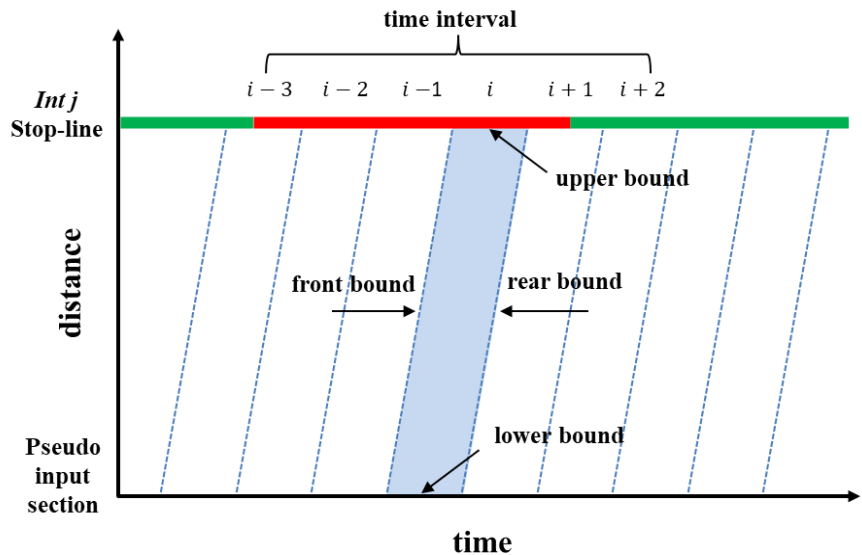

**Fig. 3.** Illustration of spatial-temporal zones

For the $i$-th space-time zone at intersection $j$, it is delineated by four lines: the front, rear, upper, and lower boundaries, computed as follows.

$$\begin{cases} x - x_{inter}^j = v_{lane}^j \times (t - t^i) \\ x - x_{inter}^j = v_{lane}^j \times (t - t^{i+1}) \\ x = x_{inter}^j \\ x = x_{inter}^{j-1} \end{cases} \quad (7)$$

Where $x$ denotes the vertical coordinate (distance) on the spatial-temporal diagram, and $t$ signifies the horizontal coordinate (time). The four equations in Eq. (7) correspond to the front, rear, upper, and lower bounds, respectively.

Thus, the queuing wave should take the form of a polyline rather than a straight line, as illustrated in Fig. 4.

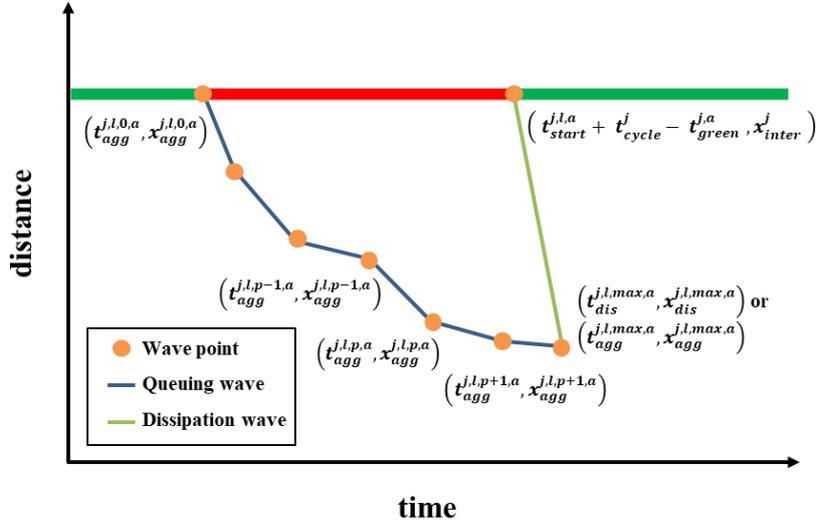

**Fig. 4.** Illustration of the queueing wave as a polyline

Referring to Fig. 4, let $A_{shockwave}^{j,l,a}$ denote a set of points defining the traffic shockwave at intersection $j$ during cycle $l$. $A_{shockwave}^{j,l,a}$ can be subdivided into two sets of points: $A_{agg}^{j,l,a}$, representing queuing shockwave points, and $A_{dis}^{j,l,a}$, representing dissipation shockwave points. This subdivision can be expressed as follows:

$$A_{dis}^{j,l,a} = (t_{dis}^{j,l,0,a}, x_{dis}^{j,l,0,a}), (t_{dis}^{j,l,1,a}, x_{dis}^{j,l,1,a})$$
$$= \{(t_{start}^{j,l,a} + t_{cycle}^j - t_{green}^{j,a}, x_{inter}^j), (t_{dis}^{j,l,max,a}, x_{dis}^{j,l,max,a})\} \quad (8)$$
$$A_{agg}^{j,l,a} = \{(t_{agg}^{j,l,0,a}, x_{agg}^{j,l,0,a}), \dots, (t_{agg}^{j,l,p,a}, x_{agg}^{j,l,p,a}), \dots, (t_{agg}^{j,l,max,a}, x_{agg}^{j,l,max,a})\} \quad (9)$$

Where $t_{start}^{j,l,a}$ represents the start time of cycle $l$ for lane $a$ at intersection $j$; $t_{cycle}^j$ denotes the cycle duration of intersection $j$; $t_{green}^{j,a}$ signifies the green light time for lane $a$ at intersection $j$; $x_{inter}^j$ stands for the vertical coordinate of the stop-line at intersection $j$; $(t_{dis}^{j,l,max,a}, x_{dis}^{j,l,max,a})$ represents the coordinates of the dissipation wave's endpoint in cycle $l$ for lane $a$ at intersection $j$; $t_{agg}^{j,l,p,a}$ is the horizontal coordinate of the point where lane a at intersection $j$ falls within the $p$-th segment of the wave in cycle $l$; $x_{agg}^{j,l,p,a}$ signifies the vertical coordinate of the point where lane $a$ at intersection $j$ falls within the $p$-th segment of the wave in cycle $l$; $(t_{agg}^{j,l,max,a}, x_{agg}^{j,l,max,a})$ denotes the coordinate of the endpoint of the queuing wave for lane $a$ at intersection $j$ in cycle $l$.

To calculate $A_{agg}^{j,l}$, convert traffic volume to traffic flow rate $q_{down}^{i,j}$, denoting the arrival flow rate at intersection $j$ during time interval $i$, as computed in Eq. (10). Subsequently, calibrate it to lane-based traffic flow rate $q_{down}^{i,j,a}$ using Eq. (11).

$$q_{down}^{i,j} = \frac{n_{down}^{i,j}}{\Delta_t} \quad (10)$$

$$q_{down}^{i,j,a} = q_{down}^{i,j} \times \beta_{lane}^{j,l,a} \quad (11)$$

Where $q_{down}^{i,j}$ is the traffic flow rate for all lanes of the specific approach; $q_{down}^{i,j,a}$ signifies the traffic flow rate for lane $a$ at intersection $j$ during time interval $i$; $\beta_{lane}^{j,l,a}$ denotes the proportion of traffic allocated to lane $a$ at intersection $j$ within that particular approach, easily calibrated using LPR data.

Thus, the velocity of queuing wave can be calculated:

$$\omega_{agg}^{i,j,l,a} = \frac{0 - q_{down}^{i,j,a}}{k_{queue} - k_{down}^{i,j,a}} \quad (12)$$

Where $\omega_{agg}^{i,j,l,a}$ denotes the queuing shockwave velocity for lane $a$ at intersection $j$ during the $i$-th spatial-temporal zone of cycle $l$; $k_{queue}$ represents the vehicle density within the queue; $k_{down}^{i,j,a}$ signifies the traffic flow density arriving at lane $a$ of intersection $j$ from the upstream in the $i$-th spatial-temporal zone, calculated based on traffic shockwave theory as follows:

$$k_{down}^{i,j,a} = \frac{q_{down}^{i,j,a}}{v_{lane}^{j}} \quad (13)$$

Where $v_{lane}^{j}$ represents the average travel speed from the pseudo input section to the stop-line of intersection $j$.

Therefore, by combining Eq. (12) and Eq. (13), we can derive the queuing wave velocity as follows:

$$\omega_{agg}^{i,j,l,a} = \frac{-q_{down}^{i,j}}{k_{queue} - q_{down}^{i,j}/v_{lane}^{j}} \quad (14)$$

Therefore, the queuing shockwave velocity for the $i$-th spatial-temporal zone can be established. Regarding cycle $l$ for lane $a$, multiple spatial-temporal zones may be encompassed, each representing an individual queuing shockwave that collectively forms the overall queuing wave for the cycle. Initially, the equation describing the queuing shockwave for each spatial-temporal zone can be computed as follows. In this manuscript, each unique spatial-temporal zone $i$ corresponds to a specific set of queuing shockwave points denoted by $p$.

$$x - x_{agg}^{j,l,p,a} = \omega_{agg}^{i,j,l,a} \times (t - t_{agg}^{j,l,p,a}) \quad (15)$$

For $t \geq t_{start}^{j,l,a}$, the intersection points $(t_{agg}^{j,l,1,a}, x_{agg}^{j,l,1,a})$ between the first segment of the polylinear queuing wave and the rear bound of the corresponding spatial-temporal zone can be computed as follows.

$$\begin{cases} t_{agg}^{j,l,1,a} = \dfrac{\omega_{agg}^{i,j,l,a} t_{start}^{j,l,a} - v_{lane}^{j} t^{i}}{\omega_{agg}^{i,j,l,a} - v_{lane}^{j}} \\ x_{agg}^{j,l,1,a} = x_{inter}^{j} + v_{lane}^{j} \times (t_{agg}^{j,l,1,a} - t^{i}) \end{cases} \quad (16)$$

Here, for $p = 0$, indicating the start-time of current cycle $l$ and the onset of the queuing wave, the coordinates $(t_{agg}^{j,l,0,a}, x_{agg}^{j,l,0,a})$ can be set as the coordinates of the cycle's onset $(t_{start}^{j,l,a}, x_{inter}^{j})$. Here, $t_{agg}^{j,l,0,a} = t_{start}^{j,l,a} \in [t^{i0-1}, t^{i0})$.

For $p \geq 1$, by computing the intersection point between the $p$-th segment of the polylinear queuing shockwave and the rear bound of the following spatial-temporal zone, we can determine the coordinates of the $(p+1)$-th segment of the queuing wave $(t_{agg}^{j,l,p+1,a}, x_{agg}^{j,l,p+1,a})$ based on the coordinates of the final segment of the queuing wave $(t_{agg}^{j,l,p,a}, x_{agg}^{j,l,p,a})$.

Therefore, the queuing shockwave for each spatial-temporal zone can be expressed as shown, and the recursive coordinate calculation process is illustrated in Fig. 5.

$$x_{agg}^{j,l,p+1,a} = \begin{cases} x_{inter}^{j}, & p = 0 \\ x_{inter}^{j} + v_{lane}^{j} \times (t_{agg}^{j,l,p+1,a} - t^{i+1}), & p > 0 \end{cases} \quad (17)$$

$$t_{agg}^{j,l,p+1,a} = \begin{cases} t_{start}^{j,l,a}, & p = 0 \\ \dfrac{x_{inter}^{j} - v_{lane}^{j} t^{i+1} - x_{agg}^{j,l,p,a} + \omega_{agg}^{i+1,j,l,a} t_{agg}^{j,l,p,a}}{\omega_{agg}^{i+1,j,l,a} - v_{lane}^{j}}, & p > 0 \end{cases} \quad (18)$$

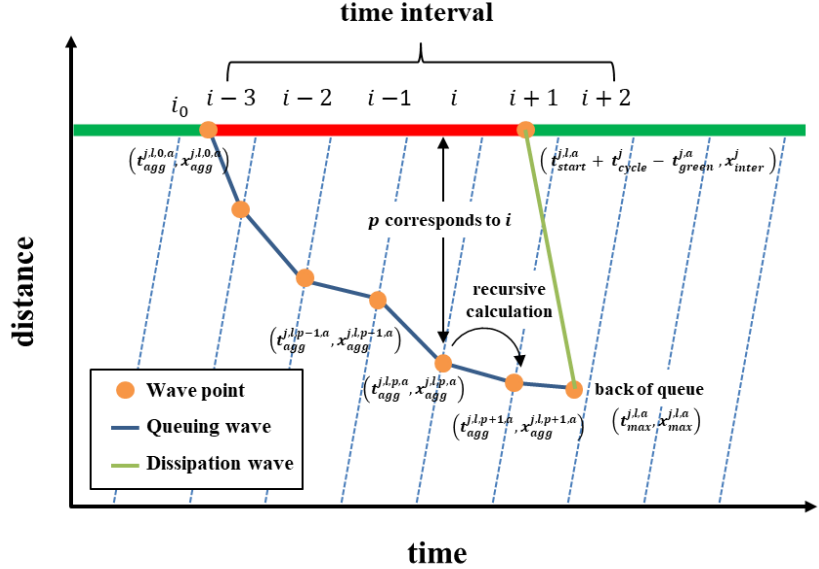

**Fig. 5.** Illustration of traffic wave points calculation

Additionally, the equation for the dissipation shockwave can be formulated as follows. In this equation, $\omega_{dis}^{j}$ denotes the dissipation wave velocity at intersection $j$.

$$x - x_{inter}^{j} = \omega_{dis}^{j} \times [t - (t_{start}^{j,l,a} + t_{cycle}^{j} - t_{green}^{j,l,a})] \quad (19)$$

The queuing shockwave can be recursively computed from the initial coordinate to construct the entire polylinear queuing shockwave. The calculation process concludes when the polylinear queuing wave intersects with the linear dissipation wave, resulting in the intersection node $(t_{max}^{j,l,a}, x_{max}^{j,l,a})$, representing the back of the queue and determined as follows.

$$\begin{cases} x_{dis}^{j,l,max,a} - x_{inter}^{j} = \omega_{dis}^{j} \times [t_{dis}^{j,l,max,a} - (t_{start}^{j,l,a} + t_{cycle}^{j} - t_{green}^{j,l,a})] \\ x_{dis}^{j,l,max,a} - x_{agg}^{j,l,p,a} = \omega_{agg}^{i,j,l,a} \times (t_{dis}^{j,l,max,a} - t_{agg}^{j,l,p,a}) \\ t_{max}^{j,l,a} = t_{agg}^{j,l,max,a} = t_{dis}^{j,l,max,a} \\ x_{max}^{j,l,a} = x_{agg}^{j,l,max,a} = x_{dis}^{j,l,max,a} \end{cases} \quad (20)$$

The coordinates of the back of queue $(t_{max}^{j,l,a}, x_{max}^{j,l,a})$ can be expressed as follows.

$$\begin{cases} t_{max}^{j,l,a} = \begin{bmatrix} x_{agg}^{j,l,p,a} - \omega_{agg}^{i,j,l,a} t_{agg}^{j,l,p,a} - x_{inter}^{j} + \omega_{dis}^{j} t_{start}^{j,l,a} \\ + \omega_{dis}^{j} (t_{cycle}^{j} - t_{green}^{j,l,a}) \end{bmatrix} \\ \qquad \times \dfrac{1}{\omega_{dis}^{j} - \omega_{agg}^{i,j,l}} \\ x_{max}^{j,l,a} = x_{dis}^{j,l,max,a} = \omega_{agg}^{i,j,l,a} \times (t_{dis}^{j,l,max,a} - t_{agg}^{j,l,p,a}) + x_{agg}^{j,l,p,a} \end{cases} \quad (21)$$

The spatial-temporal zone in which the coordinates of back of queue $(t_{max}^{j,l,a}, x_{max}^{j,l,a})$ lie can also be expressed as Eq. (22), and the corresponding spatial-temporal zone is denoted by $i_{max}^{j,l,a}$. Hence, the coordinates of all traffic wave points in $A_{shockwave}^{j,l,a}$ can be calculated, as depicted in Fig. 5.

$$\text{s.t.} \begin{cases} x_{agg}^{j,l,max,a} - x_{inter}^{j} \geq v_{lane}^{j} \times (t_{agg}^{j,l,max,a} - t^{i}) \\ x_{agg}^{j,l,max,a} - x_{inter}^{j} < v_{lane}^{j} \times (t_{agg}^{j,l,max,a} - t^{i-1}) \end{cases} \quad (22)$$

## D. Traffic Volume Estimation

After obtaining the arrival flow profile and its associated shockwave, vehicles dissipating from these outer sections contribute to the input traffic flow of the subsequent downstream intersection. Consequently, to recursively estimate the arrival flow profile and corresponding shockwave of the next downstream intersection, it is crucial to assess the dissipated traffic flow distribution within the current cycle.

Initially, using the computed coordinates of traffic wave points in $A_{shockwave}^{j,l,a}$, we can determine the timestamp of the last queued vehicle crosses the stop-line using Eq. (23). Consequently, the total number of queued vehicles can be computed based on the polylinear queuing wave using Eq. (24).

$$t_{clear}^{j,l,a} = t_{agg}^{j,l,i_{max}^{j,l,a},a} + \frac{x_{inter}^{j} - x_{agg}^{j,l,i_{max}^{j,l,a},a}}{v_{inter}^{j}} \quad (23)$$

$$n_{queue}^{j,l,a} = \sum_{i=i_{start}^{j,l,a}}^{i_{max}^{j,l,a}} n_{down}^{i,j,a} \quad (24)$$

Where $t_{clear}^{j,l,a}$ signifies the timestamp of the last queued vehicle in lane $a$ at intersection $j$ passes the stop-line during cycle $l$; $n_{queue}^{j,l,a}$ denotes the number of queued vehicles in lane $a$ of intersection $j$ in cycle $l$; $i_{start}^{j,l,a}$ denotes the serial number of the spatial-temporal zone which the onset of cycle lies.

Furthermore, the onset of the green light for lane $a$ in cycle $l$ at intersection $j$, $t_{gstart}^{j,l,a}$, can be calculated as $t_{gstart}^{j,l,a} = t_{start}^{j,l+1,a} - t_{green}^{j,a}$, and the corresponding serial number of the spatial-temporal zone can be denoted as $i_{gstart}^{j,l,a}$. Thus, the queued vehicles $n_{queue}^{j,l,a}$ that dissipate during the time interval $i_{gstart}^{j,l,a}$ to $i_{clear}^{j,l,a}$ can be considered to dissipate evenly. For the subsequent time interval, spanning from $i_{clear}^{j,l,a} + 1$ to $i_{start}^{j,l+1,a} - 1$, the dissipated vehicles are not influenced by signal control. Therefore, the dissipated traffic flow remains equal to its corresponding arrival traffic flow, denoted as $n_{down}^{i,j,a}$. Consequently, the dissipated traffic can be calculated as follows.

$$n_{inter}^{i,j,a} = \begin{cases} 0, & i \in [i_{start}^{j,l,a}, i_{gstart}^{j,l,a} - 1] \\ \dfrac{n_{queue}^{j,l,a}}{i_{clear}^{j,l,a} - i_{gstart}^{j,l,a} + 1}, & i \in [i_{gstart}^{j,l,a}, i_{clear}^{j,l,a}] \\ n_{down}^{i,j,a}, & i \in [i_{clear}^{j,l,a} + 1, i_{start}^{j,l+1,a} - 1] \end{cases} \quad (25)$$

Where $n_{inter}^{i,j,a}$ denotes the dissipated traffic volume of lane $a$ of intersection $j$ during time interval $i$.

As dissipating vehicles cross the stop-line, some will exit the arterial road. Consequently, the traffic flow towards the subsequent downstream intersection, $n_{join}^{i,j+1,th}$, can be determined as follows.

$$n_{join}^{i,j+1,th} = \sum_a n_{inter}^{i,j,a} \times (1 - \alpha_{leave}^{j,a}) \quad (26)$$

Where $n_{join}^{i,j+1,th}$ represents the traffic flow directed towards the subsequent downstream intersection $j+1$ originating from the through-ahead movement at intersection $j$; $\alpha_{leave}^{j,a}$ signifies the proportion of lane $a$ at intersection $j$ that diverts from the arterial road, easily calibrated using historical LPR data.

Assuming the dissipated traffic flow travels at a speed of $v_{inter}^{j+1,th}$ from intersection $j$ (with the horizontal coordinate denoted as $x_{inter}^{j}$) to intersection $j+1$ (with the horizontal coordinate denoted as $x_{inter}^{j} + x_{join}^{j+1,th}$), it is crucial to consider that the traffic flow has already been aggregated into time intervals of $\Delta_t$ at intersection $j$. Consequently, the sequential numbering of the aggregated traffic flow to intersection $j+1$ may not align with the sequential numbering of time intervals at intersection $j+1$. Therefore, adjusting the traffic flow becomes necessary to synchronize the spatial-temporal aspects of intersection $j+1$ with the arterial.

Initially, considering the dissipation traffic flow $n_{join}^{i_0,j+1,th}$ at any time interval $i_0$ from intersection $j$, where the start time of the time interval is $i_0 \times \Delta_t$, and the coordinate on the spatial-temporal diagram is $(i_0 \times \Delta_t, x_{inter}^{j})$. Assuming that the dissipation traffic flow travels at a speed of $v_{inter}^{j+1,th}$ from intersection $j$, the coordinate upon arrival at the subsequent intersection becomes $(i_0 \times \Delta_t + x_{join}^{j+1,th}/v_{inter}^{j+1}, x_{inter}^{j} + x_{join}^{j+1,straight})$. Subsequently, determining the time interval within which this coordinate resides can be executed as follows.

$$i_1 \times \Delta_t \leq i_0 \times \Delta_t + \frac{x_{join}^{j+1,th}}{v_{inter}^{j+1}} < (i_1 + 1) \times \Delta_t \quad (27)$$

Therefore, at the location $x_{inter}^{j} + x_{join}^{j+1,th}$, the traffic flow $n_{up}^{i_1,j+1,th}$ during the time period from $i_1 \times \Delta_t$ to $(i_1 + 1) \times \Delta_t$ can be divided into two segments: $n_{up1}^{i_1,j+1,th}$ during the time period from $i_1 \times \Delta_t$ to $i_0 \times \Delta_t + x_{join}^{j+1,th}/v_{inter}^{j+1}$ and $n_{up2}^{i_1,j+1,th}$ during the time period from $i_0 \times \Delta_t + x_{join}^{j+1,straight}/v_{inter}^{j+1}$ to $(i_1 + 1) \times \Delta_t$. Furthermore, both $n_{up1}^{i_1,j+1,th}$ and $n_{up2}^{i_1,j+1,th}$ are essentially derived from modifications of the traffic flow during time intervals $i_0$ and $i_0 - 1$. Therefore, their calculation is as follows.

$$\begin{cases} n_{up1}^{i_1,j+1,th} = n_{join}^{i_0-1,j+1,th} \times \dfrac{i_0 \times \Delta_t + \dfrac{x_{join}^{j+1,th}}{v_{inter}^{j+1}} - i_1 \times \Delta_t}{\Delta_t} \\ \\ n_{up2}^{i_1,j+1,th} = n_{join}^{i_0,j+1,th} \times \dfrac{(i_1+1) \times \Delta_t - \left(i_0 \times \Delta_t + \dfrac{x_{join}^{j+1,th}}{v_{inter}^{j+1}}\right)}{\Delta_t} \end{cases} \quad (28)$$

Afterwards, the calculation of the adjusted dissipated traffic flow to intersection $j+1$ in time interval $i$, $n_{up}^{i,j+1,th}$, can be conducted as follows.

$$n_{up}^{i_1,j+1,th} = n_{join}^{i_0-1,j+1,th} \times \dfrac{i_0 \times \Delta_t + \dfrac{x_{join}^{j+1,th}}{v_{inter}^{j+1}} - i_1 \times \Delta_t}{\Delta_t} \\ + n_{join}^{i_0,j+1,th} \times \dfrac{(i_1+1) \times \Delta_t - \left(i_0 \times \Delta_t + \dfrac{x_{join}^{j+1,th}}{v_{inter}^{j+1}}\right)}{\Delta_t} \quad (29)$$

Similarly, $n_{up}^{i,j+1,r}$ and $n_{up}^{i,j+1,l}$, representing the dissipated traffic flow of the right-turning and left-turning movements from intersection $j$, can also be determined. Consequently, the input traffic flow for link $j+1$ during time interval $i$, $n_{up}^{i,j+1}$, can be computed.

$$n_{up}^{i,j+1} = n_{up}^{i,j+1,left} + n_{up}^{i,j+1,straight} + n_{up}^{i,j+1,right} \quad (30)$$

After obtaining the input traffic volume for link $j+1$ at intersection $j+1$, the arrival flow profile of intersection $j+1$ can be estimated using the method similarly. Once the recursive estimation process for all intersections is completed, the arrival flow profile can be determined.

## IV. ARRIVAL FLOW PROFILE PREDICTION

The previous section centered on estimating the arrival flow profile under predefined arterial signal timing conditions. This section introduces a method for simulating and predicting the dynamic arrival flow profile in scenarios without predefined arterial signal timing, referred to as arrival flow profile prediction.

As the arrival flow profile is affected by changes in arterial signal timing, the observed vehicle data of LPR data may not accurately represent traffic arrival information for the outer section. Therefore, it is crucial to initially adjust the input data for all arterials. To acquire the timestamp of vehicles passing the stop-line, a Poisson distribution is used for the re-generation of timestamps for non-queued vehicles. However, this re-generated distribution may not completely adhere to the fundamental mechanisms of traffic flow operation, necessitating further adjustments to the arrival timestamps of certain vehicles.

For the re-generated arrival vehicular data, a sequential examination of headways should be conducted starting from the leading vehicle. When the headway between the front and rear vehicle is less than the saturated headway, the arrival timestamp of the rear vehicle must be adjusted to comply with the constraint. Subsequently, comprehensive information on the arrived vehicles can be compiled, corresponding to the $q_{down}^{i,j}$ as discussed in the previous section, as illustrated in Fig. 6. The estimation of the arrival flow profile can then be conducted similarly.

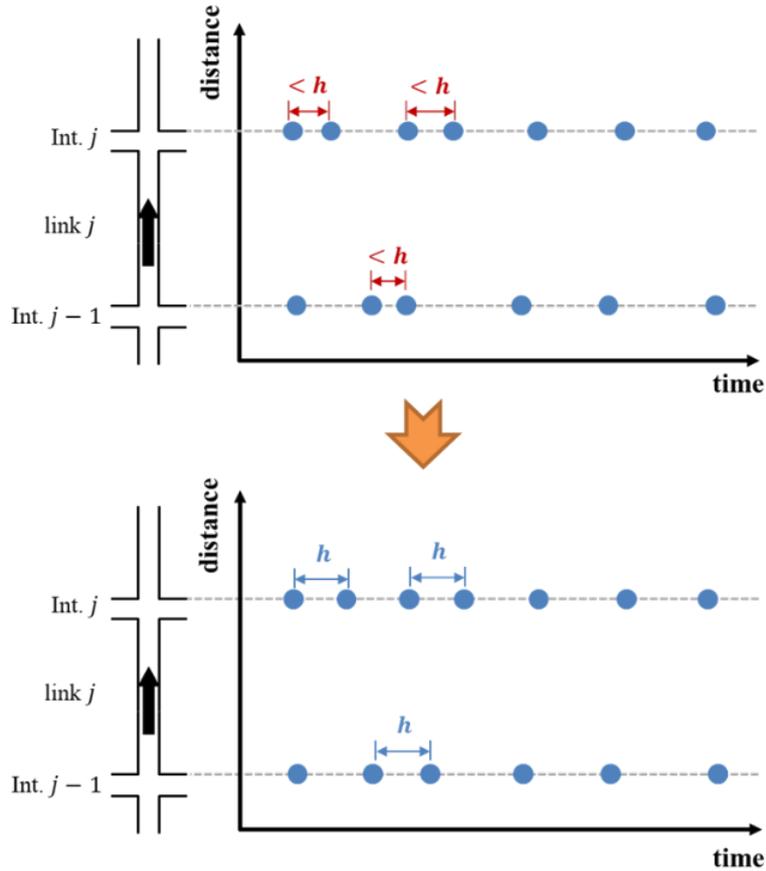

**Fig. 6.** Illustration of timestamp modification

Similarly, for the inner section, once the input traffic flow has been acquired and adjusted, predicting the arrival flow profile follows the same methodology as the arrival flow profile estimation method introduced in the preceding section.

## V. VERIFICATION

The proposed method was implemented using PYTHON and subsequently evaluated through an empirical case and a simulation case, based on five consecutive intersections along South Fuzhou Road in Qingdao City, China. The southbound of the arterial was chosen as the object, encompassing four consecutive road links as shown in Fig. 7.

Initially, an empirical case was conducted to validate the accuracy of the arrival flow profile estimation based on full-sampled LPR data, trajectory data, and signal timing data. Subsequently, to further validate the accuracy of the arrival flow profile prediction and its sensitivity to various signal timings, a simulation case was executed using the VISSIM simulation platform, aligning the input traffic flow with ground-truth data.

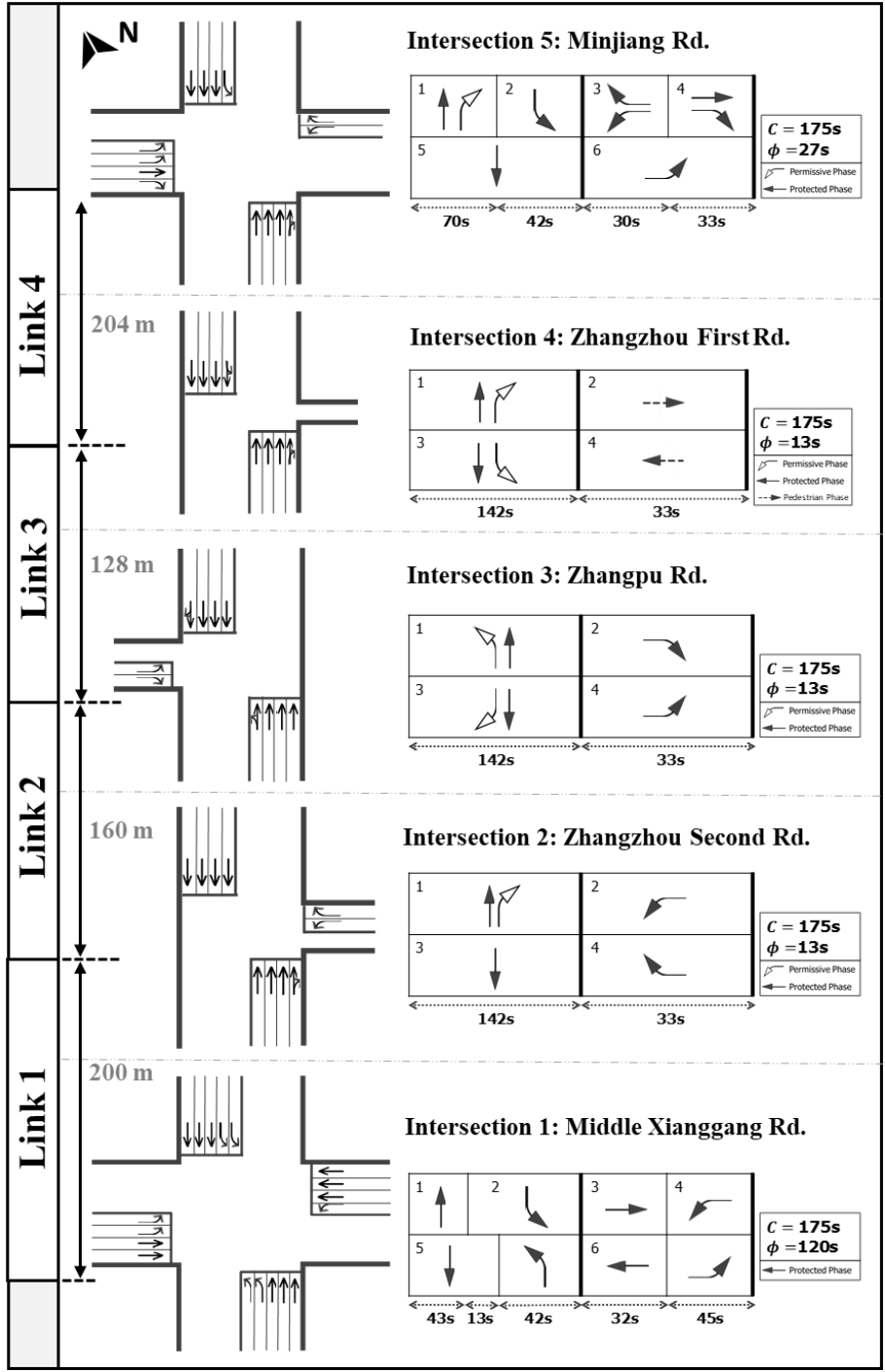

**Fig. 7.** Illustration of geometry and signal timing

## A. Verification of Arrival Flow Profile Estimation by Empirical Case

Due to the availability of data, full-sample LPR data, trajectory data, and signal timing data for links 2, 3, and 4 were extracted from video recordings. The video data from November 29, 2016, was chosen for validation, covering the time period from 07:15:00 AM to 08:15:00 AM. Real-time signal timing data was extracted and shown in Fig. 7. Additionally, simulative LPR data, mirroring the attributes of actual LPR data, was also extracted. The full-sample trajectory data was then extracted using semi-automatic software, George, enabling vehicle tracking at intervals of 0.12 seconds, as illustrated in Fig. 8 [28].

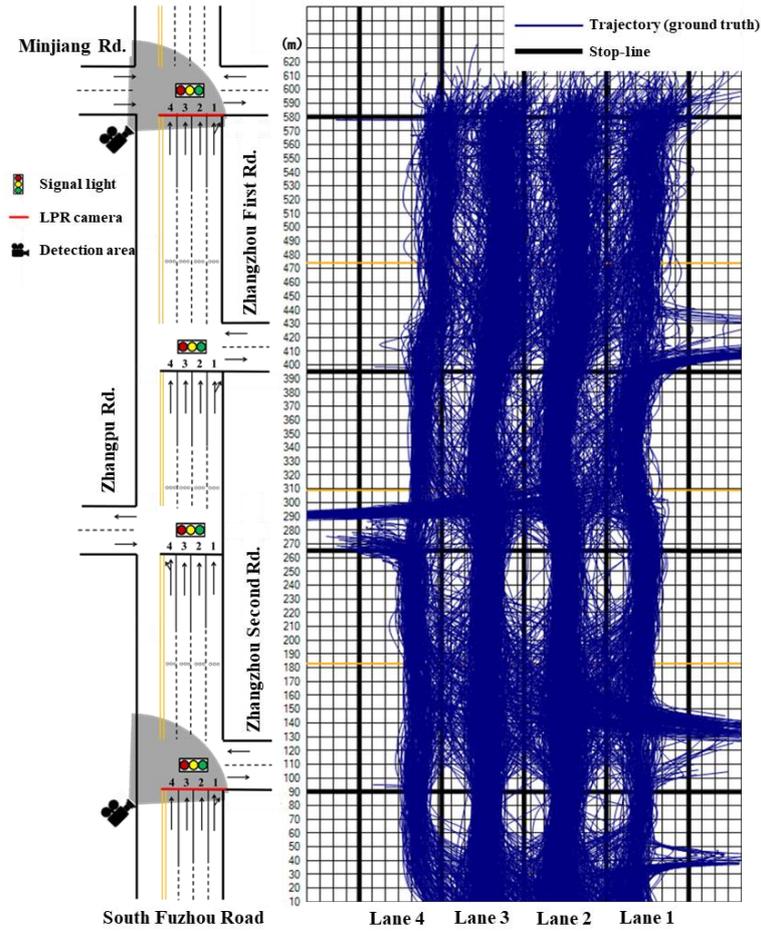

**Fig. 8.** Illustration of extracted trajectory data

Initially, the unit time interval for the verification case, denoted as $\Delta_t$, was configured to 5 seconds. Then, based on full-sampled data to verify, as for any given vehicle $i$, four indictors were defined to verify the accuracy of arrival flow profile: offset $\delta_i^1$, offset $\delta_i^2$, ratio $r_i^1$ and ratio $r_i^2$. $\delta_i^1$ indicates the vehicular delay not covered by the area bounded by the estimated arrival flow profile. $\delta_i^2$ indicates the offset between estimated arrival flow profile and the onset of vehicular delay; $r_i^1$ indicates the ratio of vehicular delay covered by the area bounded by estimated arrival flow profile; $r_i^2$ indicates the ratio of offset $\delta_i^2$ with the length on horizontal axis.

As shown in Fig. 9, indicators $\delta_i^1$ and $r_i^1$ were defined to describe the ability of depicting vehicular driving processes, while indicators $\delta_i^2$ and $r_i^2$ were defined to verify if the arrival flow profile is overestimated. A good arrival flow profile estimate should be analyzed by examining all indicators. Thus, for any given vehicle $i$, these four indicators can be calculated using Eq. (31-34), and for all time intervals of verification, these four indicators can be calculated using Eq. (35-38).

$$\delta_i^1 = d_i^{uc} \tag{31}$$
$$\delta_i^2 = t_i^{uc} \tag{32}$$
$$r_i^1 = d_i^{uc}/(d_i^{uc} + d_i^c) \tag{33}$$
$$r_i^2 = t_i^{uc}/t_i^a \tag{34}$$

Where $d_i^{uc}, d_i^c$ denotes the vehicular delay which is not covered and covered by the area bounded by the estimated arrival flow profile for vehicle $i$, respectively; $t_i^{uc}$ denotes the offset between estimated arrival flow profile and the onset of vehicular delay for vehicle $i$; $t_i^a$
16

denotes the length of the area bounded by estimated arrival flow on horizontal axis.

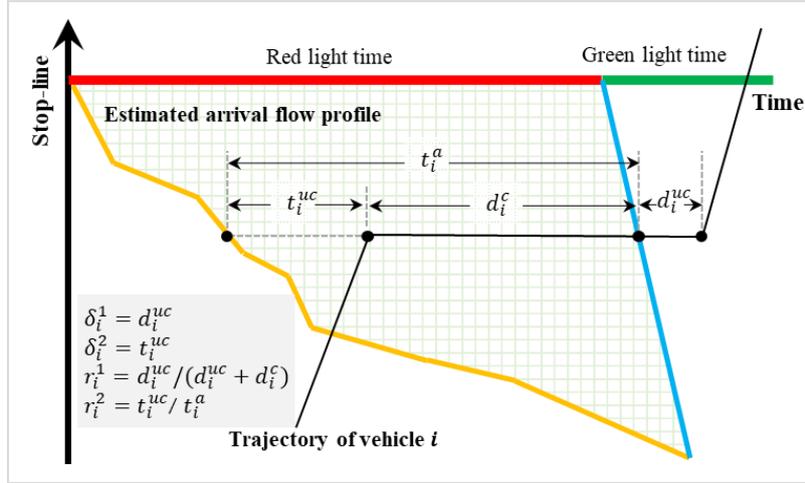

**Fig. 9.** Illustration of evaluation indicators

$$\delta^1 = \sum_{i=1}^{N} \frac{d_i^{uc}}{N} \tag{35}$$

$$\delta^2 = \sum_{i=1}^{N} \frac{t_i^{uc}}{N} \tag{36}$$

$$r^1 = \sum_{i=1}^{N} \frac{d_i^{uc}}{(d_i^{uc} + d_i^c) * N} \tag{37}$$

$$r^2 = \sum_{i=1}^{N} \frac{t_i^{uc}}{t_i^a * N} \tag{38}$$

Where $\delta^1, \delta^2, r^1, r^2$ denote the average value of defined four indicators over the entire period; $N$ denotes the number of vehicles (trajectory) in the period.

TABLE 1 presents the validation results, with indicator $\delta^1$ yielding values of 0.27s, 0.96s, and 5.18s for the three respective links. Concurrently, indicator $r^1$ exhibit values of 97.55%, 93.89%, and 90.01% for the same links. The combination of $\delta^1$ and $r^1$ underscores the precise representation of the dynamic queuing process by the estimated arrival flow profile. On average, the estimated profile covers 93.82% of vehicular delay, with an average uncovered delay offset of merely 2.14s. Besides, the results for indicator $\delta^2$ in the three links are 2.71s, 1.34s, and 16.53s, and the corresponding indicator $r^2$ show values of 10.04%, 8.55%, and 19.56%, respectively. This highlights the proximity between the estimated arrival flow profile and the actual queuing wave. Despite a larger offset in link 3 (16.53s), the relative error remains within an acceptable range at 19.56%.

**TABLE 1 Validation results of arrival flow profile estimation**

| Indicator | Link 2 | Link 3 | Link 4 | Average |
| --- | --- | --- | --- | --- |
| $\delta^1$ | 0.27 s | 0.96 s | 5.18 s | 2.14 s |
| $\delta^2$ | 2.71 s | 1.34 s | 16.53 s | 6.86 s |
| $r^1$ | 97.55 % | 93.89 % | 90.01 % | 93.82% |
| $r^2$ | 10.04 % | 8.55 % | 19.56 % | 12.72% |

Although the verification results for three links are promising, the results for link 4 are comparatively lower than the others. The reason is that the signal timing of intersections comprising link 2 and link 3 was coordinated, resulting in slightly estimated arrival flow profiles and mostly non-queued trajectories, ensuring promising accuracy. However, for link 4, the lack of coordination between intersections and a substantial deviation in the distance between queued vehicles made it challenging to calibrate as a fixed queued density value, resulting in an increased error between the arrival flow profile and trajectory.

For a more visually informative representation of the estimation results, the time interval between 07:40:00 and 08:00:50 was selected for illustration, as depicted in Fig. 10. The horizontal axis represents time, and the vertical axis corresponds to the arterial distance. The orange polygons outline the arrival flow profile for each road link, while the light green bars indicate the dissipation of traffic flow towards the subsequent road link along the arterial. The black line represents the actual extracted vehicular trajectory.

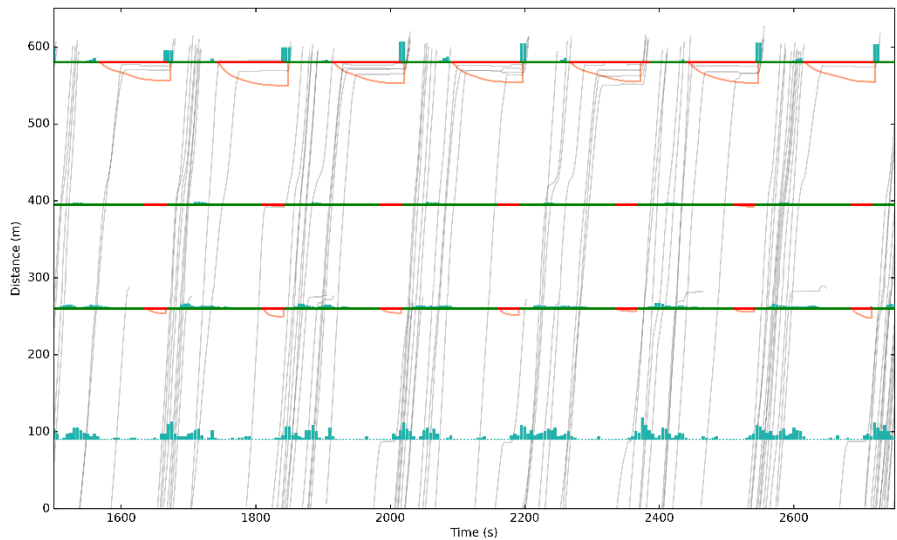

**Fig. 10.** Illustration of arrival flow profile estimation

From Fig. 10, it is evident that the coordinated signal timing scheme has led to a subtle arrival flow profile, with link 4 displaying a more obvious profile. The region enclosed by the estimated arrival flow profile effectively illustrates the dynamic spatial-temporal progression of vehicles. It is noteworthy that a considerable portion of the traffic volume exits midway along the link, rather than at the intersection. Based on original video, it becomes apparent that points of intersects, such as residential communities and parking lots, contribute to potential errors in the subsequent road link analysis.

**B. Verification of Arrival Flow Profile Prediction by Simulation Case**

To assess the accuracy of arrival flow profile prediction in the absence of predefined arterial signal timing, a simulation case, incorporating all four links, was conducted using the VISSIM simulation platform. The simulation model underwent calibration using actual geometric and traffic flow data. In VISSIM, data collection detectors were strategically placed at the stop-lines of each lane to mimic real-world LPR cameras as input. The information collected by these detectors corresponds precisely to actual LPR data.

To validate the accuracy of arrival flow profile prediction under varying signal timing schemes, demand-to-capacity (D/C) ratio and effective bandwidth (EB) ratio were considered as verification factors. The D/C ratio was related to the green split of each intersection, with values of 0.4, 0.6, and 0.8 chosen for verification. The EB ratio signifies the ratio between the coordination bandwidth with the green split of the target intersection, linked to coordination offset, with values of 100%, 50%, and 0% chosen for verification, where 100% indicates an optimal coordination scheme, and 0% denotes the least favorable condition. Thus, considering these two factors, crossover experiments were designed, comprising nine cases detailed in TABLE 2.

For each case in TABLE 2, the simulation duration was set at 4200 seconds, encompassing a warm-up period of 600 seconds. The random seed for each case was uniformly set to 12. The unit time interval $\Delta_t$ for the verification case was established as 5 seconds, resulting in a total of 720 data points to depict the arrival flow profile for each road link throughout the study duration. Additionally, four defined indicators, calculable using Equations (31-38), were chosen for verification, and the validation results are presented in TABLE 2.

**TABLE 2 validation results of arrival flow profile prediction**

| Case No. | D/C ratio | EB ratio | $\delta^1$ | $\delta^2$ | $r^1$ | $r^2$ |
|---|---|---|---|---|---|---|
| 1 | 0.4 | 100% | 1.06 s | 4.23 s | 90.05% | 18.46% |
| 2 | 0.4 | 50% | 0.51 s | 7.35 s | 93.31% | 31.32% |
| 3 | 0.4 | 0% | 0.44 s | 5.81 s | 94.11% | 25.15% |
| 4 | 0.6 | 100% | 2.40 s | 7.33 s | 85.75% | 15.47% |
| 5 | 0.6 | 50% | 0.60 s | 10.42 s | 90.99% | 20.58% |
| 6 | 0.6 | 0% | 1.47 s | 11.65 s | 93.46% | 26.62% |
| 7 | 0.8 | 100% | 5.05 s | 4.81 s | 82.28% | 11.56% |
| 8 | 0.8 | 50% | 3.74 s | 9.90 s | 81.12% | 21.74% |
| 9 | 0.8 | 0% | 1.75 s | 12.21 s | 84.21% | 16.56% |
| Average | | | 1.89 s | 8.19 s | 88.36% | 20.83% |

TABLE 2 presents the results of arrival flow profile prediction for the entire arterial network, comprising four road links. The overall average value of indicator $\delta^1$ is 1.89 s, and the overall average value of indicator $r^1$ is 88.36%. These values signify that the area bounded by the estimated arrival flow profile precisely captures the spatial-temporal queuing process, leaving only a 1.89s vehicular delay uncovered while covering 88.36% of vehicular delay. Additionally, the overall average value of $\delta^2$ is 8.19 s, with an overall average value of $r^2$ at 20.83%. This indicates that the estimated arrival flow profile closely aligns with the queuing shockwave, exhibiting an acceptable relative error of 20.83%.

It is noteworthy that for indicators $\delta^1$ and $r^1$, representing the ability to depict the queuing process, there is a negative relationship with the EB ratio under the same D/C ratio (i.e., case1, case2, case3). In most instances, the optimal result is achieved when the EB ratio is minimized. This is because a decrease in coordination bandwidth enhances the significance of the arrival flow profile, effectively improving the ability to capture the spatial-temporal queuing process. However, this may also increase the error with the actual queuing wave, as evidenced by indicators $\delta^2$ and $r^2$, which exhibit a positive relationship with the EB ratio.

Furthermore, for the same EB ratio, comparing cases with different D/C ratios, such as case 1, case 4, and case 7, reveals that the accuracy of indicators $\delta^1$ and $r^1$ has a negative relationship with the D/C ratio. This implies that the accuracy of depicting the queuing process decreases with an increase in traffic demand. However, even when the D/C ratio reaches 0.8, more than 80% of vehicular delay can still be covered by the area bounded by the predicted arrival flow profile, which is considered acceptable. Additionally, the accuracy of indicators $\delta^2$ and $r^2$ shows a positive relationship with the D/C ratio, suggesting that the arrival flow profile becomes closer to the actual queuing wave as the number of vehicles increases, potentially providing more sampled data for modeling.

For a more visually informative presentation of the prediction results, Fig. 11 illustrates the predicted arrival flow profile of the arterial for case 7 (D/C ratio: 0.8, EB ratio: 100%). The elements and meanings conveyed in Fig. 11 are consistent with Fig. 10. Thus, the light green bar at the first intersection (located at a distance of 100 meters) represents the recorded input traffic flow on the arterial. Meanwhile, the orange line polygon at the intersections illustrates the shockwave profile bounded by the predicted arrival flow profile from the upstream.

Additionally, the light green bars at the remaining four intersections represent the dissipation traffic flow contributing to the formation of the arrival flow profile at the subsequent downstream intersection, divided into two parts: saturated dissipation part and unsaturated dissipation part. The noticeable unsaturated dissipation part indicates a more favorable coordination condition (e.g., intersection 5).

It is important to note that the arrival flow profile at intersections 3 and 4 is steep, potentially influenced by the D/C ratio and bandwidth. In this case, the D/C ratio is 0.8, and most cycles are over-saturated. For these cycles, a small portion of vehicles stopped multiple times. Furthermore, due to the EB ratio being 100%, implying that most arrival vehicles from upstream approach downstream around the onset of the green light time, these two processes may contribute to the steep arrival flow profile, with a piecewise relation with roughly two segments. Consequently, the proposed prediction method provides a more detailed depiction. For example, the aforementioned profile with piecewise relation is challenging to represent using conventional fixed arrival flow rates.

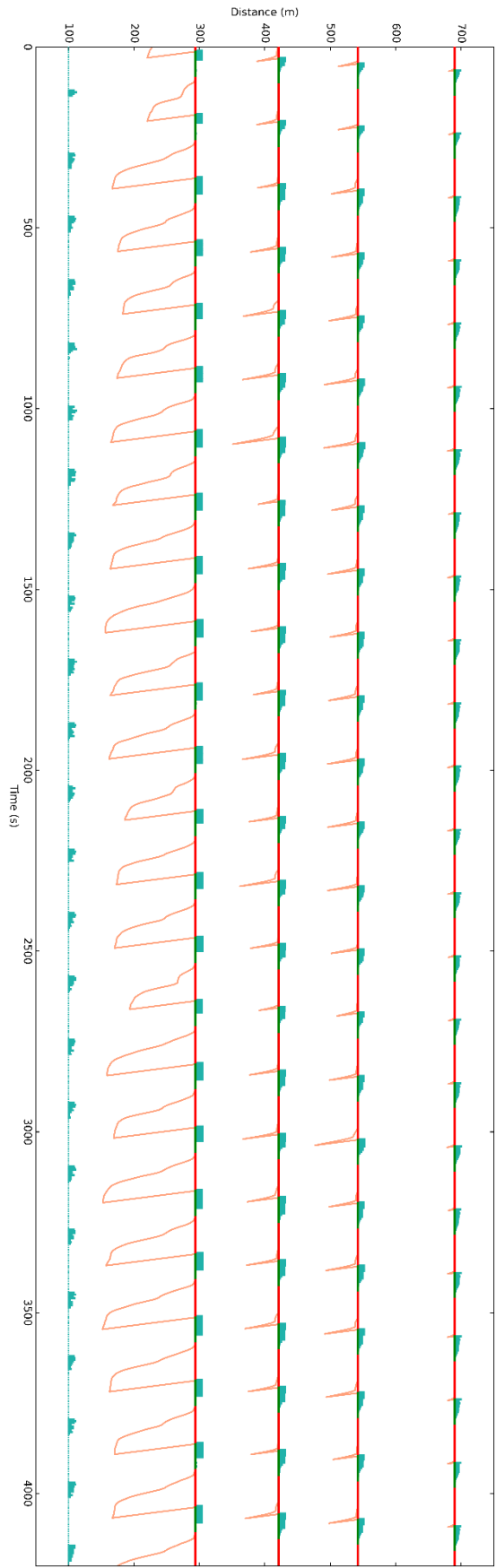

**Fig. 11.** Illustration of arrival flow profile prediction

## VI. CONCLUSION AND FUTURE WORK

In this study, we introduced novel methods for estimating and predicting arrival flow profiles in urban arterials, utilizing the wealth of information provided by LPR data. Our approach commenced with the implementation of a platoon dispersion model to calculate the arrival traffic flow within the urban arterial. Subsequently, we employed shockwave theory to achieve precise estimation of the arrival flow profile. This estimation process involved recursive calculations spanning all arterials, ensuring a comprehensive representation of traffic dynamics. Furthermore, we proposed an adjustment to the arrival vehicular information to facilitate the prediction of arrival flow profiles given any timing schemes.

Empirical validation underscores the effectiveness of our estimation method, attaining an impressive accuracy rate of 93.82% for depicting the spatial-temporal queuing process of vehicles, and an acceptable relative error of 12.72% with the actual queuing wave. Simulation case also indicates the effectiveness of arrival flow profile predication with un-predefined signal timing, with an accuracy rate of 88.36% for depicting the spatial-temporal queuing process of vehicles and the error is only 1.89 s, and the relative error of 20.83% with the actual queuing wave is also acceptable, and this predication method is feasible and can achieve a promising result under various demand-to-capacity ratio and coordination bandwidth setting. These results validate the promise of our approach in accurately estimating and predicting arrival flow profiles, providing valuable insights into urban arterial traffic conditions.

This study serves as a valuable tool for pre-implementation assessment of signal timing efficacy on urban arterials, providing a practical alternative to simulation tools and demonstrating broader applicability for signal control evaluation and optimization compared to simulation-driven approaches. The methodology opens avenues for further research, including the calculation of various indicators (e.g., queue length and delay), signal timing optimization, and trajectory reconstruction. The method exclusively relies on LPR data from the boundaries of the urban arterial, ensuring robust applicability. Moreover, it can be extended to alternative data sources with operational characteristics similar to LPR cameras, such as Radio-Frequency Identification Data (RFID) and high-resolution radar data, or expanded to aggregated traffic volume data from traditional fixed-location detectors.

However, the proposed method has the following limitations and requires future research.
- In the research scenario, the utilization of LPR cameras in all lanes of the outer section is essential for accurate arrival flow profile estimation. However, in practical cases, certain lanes may lack LPR cameras or have malfunctioning cameras. Ensuring the robustness of the proposed method in such scenarios is a crucial avenue for our future research.
- As the proposed method can predict the arrival flow profile under variable and non-predefined arterial signal timing conditions, there is potential for optimizing arterial signal timing using a similar methodology. Exploring the analytical relationship between signal timing parameters and arrival flow profiles is an avenue worthy of further investigation.

## ACKNOWLEDGEMENT

This study was greatly supported by the Innovation and Technology Fund of HKSAR (Grant No. MHP/038/23), the National Key Research & Development Program of China (Grant No. 2023YFE0209300) and the National Natural Science Foundation of China Project (Grant No. 52372319). The authors would like to thank Prof. Hideki Nakamura from Nagoya University for providing the video analysis software.

# REFERENCES

[1] G. Yang, Z. Tian, Z. Wang, H. Xu, and R. Yue, "Impact of on-ramp traffic flow arrival profile on queue length at metered on-ramps," Journal of Transportation Engineering, Part A: Systems, vol. 145, no. 2, 2019. doi:10.1061/jtepbs.0000211

[2] G. Yang, R. Yue, Z. Tian, and H. Xu, "Modeling the impacts of traffic flow arrival profiles on ramp metering queues," *Transportation Research Record: Journal of the Transportation Research Board*, vol. 2672, no. 15, pp. 85–92, 2018. doi:10.1177/0361198118782253

[3] N. Geroliminis and A. Skabardonis, "Prediction of arrival profiles and queue lengths along signalized arterials by using a Markov decision process," Transportation Research Record: Journal of the Transportation Research Board, vol. 1934, no. 1, pp. 116–124, 2005. doi:10.1177/0361198105193400112

[4] A. Sharma, D. M. Bullock, and J. A. Bonneson, "Input–output and hybrid techniques for real-time prediction of delay and maximum queue length at signalized intersections," Transportation Research Record: Journal of the Transportation Research Board, vol. 2035, no. 1, pp. 69–80, 2007. doi:10.3141/2035-08

[5] G. Vigos, M. Papageorgiou, and Y. Wang, "Real-time estimation of vehicle-count within signalized links," Transportation Research Part C: Emerging Technologies, vol. 16, no. 1, pp. 18–35, 2008. doi:10.1016/j.trc.2007.06.002

[6] Z. Amini, R. Pedarsani, A. Skabardonis, and P. Varaiya, "Queue-length estimation using real-time traffic data," 2016 IEEE 19th International Conference on Intelligent Transportation Systems (ITSC), 2016. doi:10.1109/itsc.2016.7795752

[7] H. Wu, J. Yao, k. Tang, "A Queue Length Distribution Estimation Method for Signalized Intersections Using Multi-section License Plate Recognition Data," Proceeding on the 2023 International Symposium on Transportation Data and Modelling, Ispra, Italy, 2023.

[8] X. Wu and H. X. Liu, "A shockwave profile model for traffic flow on congested urban arterials," Transportation Research Part B: Methodological, vol. 45, no. 10, pp. 1768–1786, 2011. doi:10.1016/j.trb.2011.07.013

[9] M. Ramezani and N. Geroliminis, "Queue profile estimation in congested urban networks with Probe Data," Computer-Aided Civil and Infrastructure Engineering, vol. 30, no. 6, pp. 414–432, 2014. doi:10.1111/mice.12095

[10] Y. Liu, J. Guo, and Y. Wang, "Vertical and horizontal queue models for oversaturated signal intersections with quasi-real-time reconstruction of deterministic and shockwave queueing profiles using limited mobile sensing data," Journal of Advanced Transportation, vol. 2018, pp. 1–19, 2018. doi:10.1155/2018/6986198

[11] Z. Wang, L. Zhu, B. Ran, and H. Jiang, "Queue profile estimation at a signalized intersection by exploiting the spatiotemporal propagation of shockwaves," Transportation Research Part B: Methodological, vol. 141, pp. 59–71, 2020. doi:10.1016/j.trb.2020.08.009

[12] S. Hu et al., "High time-resolution queue profile estimation at signalized intersections based on extended Kalman filtering," IEEE Transactions on Intelligent Transportation Systems, vol. 23, no. 11, pp. 21274–21290, 2022. doi:10.1109/tits.2022.3173515

[13]	Z. Liu, K. Li, and Y. Ni, "Prediction of arrival flow profile of transit stream on link," Transportation Research Procedia, vol. 25, pp. 1213–1226, 2017. doi:10.1016/j.trpro.2017.05.139
[14]	C. An, X. Guo, R. Hong, Z. Lu, and J. Xia, "Lane-based traffic arrival pattern estimation using license plate recognition data," IEEE Intelligent Transportation Systems Magazine, vol. 14, no. 4, pp. 133–144, 2022. doi:10.1109/mits.2021.3051489
[15]	C. M. Day and D. M. Bullock, "Calibration of platoon dispersion model with high-resolution Signal Event Data," Transportation Research Record: Journal of the Transportation Research Board, vol. 2311, no. 1, pp. 16–28, 2012. doi:10.3141/2311-02
[16]	L. Shen, R. Liu, Z. Yao, W. Wu, and H. Yang, "Development of dynamic platoon dispersion models for predictive traffic signal control," IEEE Transactions on Intelligent Transportation Systems, vol. 20, no. 2, pp. 431–440, 2019. doi:10.1109/tits.2018.2815182
[17]	Z. Yao et al., "An efficient heterogeneous platoon dispersion model for real-time Traffic Signal Control," Physica A: Statistical Mechanics and its Applications, vol. 539, p. 122982, 2020. doi:10.1016/j.physa.2019.122982
[18]	J. W. Van Lint, "Empirical evaluation of new robust travel time estimation algorithms," Transportation Research Record: Journal of the Transportation Research Board, vol. 2160, no. 1, pp. 50–59, 2010. doi:10.3141/2160-06
[19]	J. W. C. Van Lint and S. P. Hoogendoorn, "A robust and efficient method for fusing heterogeneous data from traffic sensors on Freeways," Computer-Aided Civil and Infrastructure Engineering, vol. 25, no. 8, pp. 596–612, 2009. doi:10.1111/j.1467-8667.2009.00617.x
[20]	X. Xie, H. van Lint, and A. Verbraeck, "A generic data assimilation framework for vehicle trajectory reconstruction on signalized urban arterials using particle filters," Transportation Research Part C: Emerging Technologies, vol. 92, pp. 364–391, 2018. doi:10.1016/j.trc.2018.05.009
[21]	K. Tang, T. Xu, A. Pan, et al., "Signal timing and detector data-based reconstruction of vehicle trajectories on urban arterials," Journal of Tongji University(Natural Science), vol. 44, no. 10, pp. 1545-1552, 2016. (in Chinese).
[22]	Z. Sun and X. (Jeff) Ban, "Vehicle trajectory reconstruction for signalized intersections using mobile traffic sensors," Transportation Research Part C: Emerging Technologies, vol. 36, pp. 268–283, 2013. doi:10.1016/j.trc.2013.09.002
[23]	N. Wan, A. Vahidi, and A. Luckow, "Reconstructing maximum likelihood trajectory of probe vehicles between sparse updates," Transportation Research Part C: Emerging Technologies, vol. 65, pp. 16–30, 2016. doi:10.1016/j.trc.2016.01.010
[24]	X. Shan et al., "Vehicle energy/emissions estimation based on vehicle trajectory reconstruction using sparse mobile sensor data," IEEE Transactions on Intelligent Transportation Systems, vol. 20, no. 2, pp. 716–726, 2019. doi:10.1109/tits.2018.2826571
[25]	L. Wei, Y. Wang, and P. Chen, "A particle filter-based approach for vehicle trajectory reconstruction using sparse probe data," IEEE Transactions on Intelligent Transportation Systems, vol. 22, no. 5, pp. 2878–2890, 2021. doi:10.1109/tits.2020.2976671

[26] X. Chen et al.,"Integrated macro-micro modelling for individual vehicle trajectory reconstruction using fixed and mobile sensor data,"Transportation Research Part C: Emerging Technologies, vol. 145, p. 103929, 2022. doi:10.1016/j.trc.2022.103929

[27] Traffic Police Detachment of Shanghai Municipal Bureau of Public Security.(2024, 1 11). Comprehensive Traffic Safety Service Management Platform. [Online]. Avaliable: https://sh.122.gov.cn/. Accessed July.25, 2023. (in Chinese).

[28] K. Suzuki, H. Nakamura,"TrafficAnalyzer-the integrated video image processing system for traffic flow analysis,"proceedings of the 13th ITS World Congress, London, England, 2006, pp. 8-12.